\documentstyle{l-aa}

\begin{document}

   \thesaurus{04            
		(04.03.1;   
		 08.01.1;   
		 08.01.3;   
		 08.06.3)}  

   \title{A catalogue of [Fe/H] determinations : 1996 edition}

   \author{
G. Cayrel de Strobel \inst{1} \and C. Soubiran \inst{2} 
\and E.D. Friel \inst{3} \and N. Ralite \inst{2} \and P. Fran\c cois \inst{1} 
}

\offprints{C. Soubiran}

\institute{
Observatoire de Paris-Meudon, DASGAL, 92195 Meudon Cedex, France
\and 
Observatoire de Bordeaux, BP 89, 33270 Floirac, France
\and 
National Science Foundation, Washington DC, USA}

\date{Received November 14 ; accepted November 18, 1996 }

   \maketitle

   \begin{abstract}

A fifth Edition of the Catalogue of [Fe/H] determinations is presented herewith. 
It contains 5946 determinations
for 3247 stars, including 751 stars in 84 associations, clusters or galaxies. The literature is 
complete up to December 1995. The 700 bibliographical references correspond to 
[Fe/H] determinations obtained from high resolution spectroscopic observations and
detailed analyses, most of them carried out with the help of model-atmospheres. The
Catalogue is made up of three formatted files :\\
File 1 : field stars\\
File 2 : stars in galactic associations and clusters, and stars in
SMC, LMC, M33\\
File 3 : numbered list of bibliographical references

The three files are only available in electronic form at the Centre de Donn\'ees Stellaires in
Strasbourg, via anonymous ftp
to cdsarc.u-strasbg.fr (130.79.128.5), or via http://cdsweb.u-strasbg.fr/Abstract.html.


   \end{abstract}


\section{Introduction}

Our knowledge of the chemical history of the Galaxy is based primarily 
on metal/hydrogen
determinations in stars. There are two ways of obtaining such determinations : by 
photometry or by spectroscopy. 
If photometric determinations 
are less time consuming and have good internal precision, 
spectroscopic detailed analysis is the only primary method from which photometry
can be calibrated. The general approach of a spectroscopic detailed analysis is to derive
metal abundances by
matching equivalent widths of weak lines of an observed spectrum to those computed 
from a grid of model
atmospheres of various effective temperatures, gravities and metallicities. 

 A first list of [Fe/H] determinations was published by Cayrel \& Cayrel de
Strobel (1966), but it was 
Pr. Hauck's idea to publish a
"Metal Abundance" Catalogue as an Appendix to the proceedings of 
the IAU symposium 72
on "Abundance effects on classification"
(Morel et al. 1976). 
 Since then, thanks to the energy of Pr. Hauck,
 four editions of the Catalogue of [Fe/H] determinations have been published 
in the A\&A Supplements (Cayrel de
Strobel et al. 1980, 1981, 1985, 1992).

 Since the end of 1970, spectroscopic abundance researches rely more and more
on high resolution, high S/N spectra taken with solid state detectors. The
increase in accuracy, thanks to an important gain in S/N ratio and a better control
of systematic errors, has sometimes reached a full order of magnitude (Cayrel 1988).
The number of stars subjected to detailed analysis has been increasing continuously
in the successive editions, as shown in Table 1.
The present version of the [Fe/H] Catalogue includes 2694 new measurements of
[Fe/H], and 1571 new stars, corresponding to an increase of 83\% and 94\% with
respect to the 1991-Edition. This strong increase in number of stars and in number
of [Fe/H] determinations does not correspond only to the use of Reticons
and CCDs in spectroscopy, but also to the development of powerful workstations and
elaborate software for data reduction, which make it possible to analyse a large
number of spectra in a reasonable time. For instance, 
three papers are included in this new Edition with
respectivly 671 stars (McWilliam 1990), 199 stars (Balachandran 1990), 183 stars 
(Edvardsson et al. 1993).

 Literature searches for references presenting 
[Fe/H] determinations from high resolution
spectroscopy were made easier than in previous versions of the Catalogue
thanks to the use of the NASA Astrophysics Data System Abstract Service. 
 The version of the Catalogue presented here should not omit
any article that has been published in the principal refereed journals 
of astrophysics before December 1995.


   \begin{table}
      \caption[]{Growth of data of the [Fe/H] Catalogue}
      \begin{flushleft}
      \begin{tabular}{lll}
            \hline
            \noalign{\smallskip}
            Reference &  Number   & Number \\
                      &  of stars & of [Fe/H] \\
            \noalign{\smallskip}
            \hline
            \noalign{\smallskip}
Cayrel \& Cayrel de Strobel 1966 & 154 & 204 \\
Morel et al. 1976 & 515 & 973 \\
Cayrel de Strobel et al. 1980 & 628 & 1109 \\
Cayrel de Strobel et al. 1981 & 707 & 1298 \\
Cayrel de Strobel et al. 1985 & 1035 & 1921 \\
Cayrel de Strobel et al. 1992 & 1676 & 3252 \\
Cayrel de Strobel et al. 1997 & 3247 & 5946 \\
            \noalign{\smallskip}
            \hline
         \end{tabular}
         \end{flushleft}
   \end{table}

The main aspects of a detailed analysis are briefly reviewed in Sect. 2.
 The presentation of the Catalogue, which was completely revised and reformatted, is
fully described in Sect. 3. Some important comments about the Catalogue are given in
Sect. 4 for the input data, and in Sect. 5 for the stellar content.
The conclusion is given in Sect. 6.


\section{Spectral detailed analysis}
To a great extent, the spectra of the stars listed in the Catalogue have been interpreted
by differential curve of growth analysis, the Sun having been adopted in the majority
of the cases as reference star. In the mid-sixties, the availability of powerful computers has pushed 
the authors to perform detailed analysis instead of coarse analysis. The theoretical line 
equivalent widths and curves of growth were then computed from detailed model atmospheres, 
interpolated in a grid of constant flux models. Authors interested in early type stars (early F,
A, Am, Ap) ordinarily use the Kurucz grid of Atlas models (Kurucz 1991a, 1991b, 1993), those interested
in late F, G and K stars use mostly the Gustafsson's grid of atmosphere models (Gustafsson et
al. 1975, Edvardsson et al. 1993). Both, Kurucz and Gustafsson models were computed
with the assumption of Local Thermodynamical Equilibrium (LTE). In a strictly differential analysis, 
this is not very important, because a large part of Non-LTE effects cancel out when the
programme star is sufficiently similar to the reference star. 
 
 A very important step in a spectral analysis of a star is the determination of accurate
atmospheric parameters for the selection of the appropriate model atmosphere. No abundance
can be derived unless the three physical parameters, effective temperature (Teff), surface gravity
(log\,g),
and microturbulence ($\xi_t$) have been obtained. The effective temperature of a star, which is the 
critical parameter, is most often derived from wide or narrow-band photometry, but in
some cases, it is derived on purely spectroscopic grounds from the comparison between $H_{\alpha}$
observed profiles and $H_{\alpha}$ computed profiles. The spectroscopic surface gravity
is determined from ionisation and excitation equilibria, as obtained from neutral and
ionized lines carefully chosen in the stellar spectrum and from wings of strong lines, broadened
by collisional damping. The microturbulence is determined by fitting a theoritical curve of growth,
 computed with an appropriate model, to the observational curve of growth. The metal abundance is 
obtained by comparing the observed equivalent widths of weak and medium-weak lines (possibily between 10
and 50 m\AA), in the spectrum of a programme star to those resulting from a model atmosphere computed
with the appropriate stellar parameters (Teff, log\,g, $\xi_t$) previously determined for the star
under investigation. Attention has been paid by the authors of the recent analyses to treat the
reference star in the same way as the programme star. According to our own experience, many
differential analyses with respect to the Sun show spurious effects just because the Sun
has not been analysed with the same grid of model atmospheres as the programme star (Cayrel de Strobel 
1996).

\section{Description of the Catalogue}

The Catalogue presents, as metal/hydrogen parameter, the usual logarithmic difference between
the relative abundance of iron in the atmosphere of a star and the relative abundance of
iron in the atmosphere of a reference star. This difference is written in the form : \\

$[Fe/H]= \log (Fe/H)_{star} - \log (Fe/H)_{ref}$ \\

Files 1 and 2 of the Catalogue contain the following columns : \\

   \begin{enumerate}
      \item {\it Identifier}\\
Great care has been taken to choose for each star its most common designation.
For field stars, the HD number was taken, when available. If not, the BD or CD or CPD number
was taken, 
with the usual convention of declinations. For very few stars only Giclas or other names
were available.
 For cluster 
stars, a great variety of
names were used in the references for the same object. 
The Dictionary of Nomenclature of Celestial Objects (Lortet et al. 1994) was consulted,
together with the SIMBAD database
and the NASA-ADS service in order to adopt the most appropriate name for each star. Except for
a dozen unidentified objects (designated with "?" at the beginning of their name), the chosen
identifier can be used in a SIMBAD interrogation. There is no
redundancy between the list of field stars and the list of cluster stars.\\

\item {\it Spectral type}\\
The spectral types come from the cross-identification of the Catalogue and 
the SIMBAD database. The same syntax has been used (see the SIMBAD user's guide 
and reference manual, chapter 15). It helps the user of the Catalogue
to locate a normal star in the HR diagram, and also to recognize peculiar stars. 
We have not corrected the spectral types, even if in a few
cases (mainly metal deficient population II stars) they disagree
with the effective temperature and gravity resulting from a detailed analysis. \\

\item {\it Object type}\\                                
This column indicates a star's peculiarity.
We have adopted the same designation as in SIMBAD database, reduced 
to two characters (see the SIMBAD user's guide and reference manual, 
appendix F). The most frequent types of peculiarities are high proper-motion
stars (PM), spectroscopic binaries (SB), variable stars (V),
stars in multiple systems (**), etc ... \\

\item {\it Visual magnitude V}\\
The sources of the visual magnitudes in SIMBAD are various and heterogeneous. As a 
consequence, this value should be considered only as an indicator of
brightness. For precise photometry, the users have to consult
a specialised catalog, like the latest edition of the General Catalogue of Photometric Data (GCPD)
(Mermilliod et al. 1996). In some cases, the magnitude indicated in this column is
the B magnitude (a letter B follows the value of the magnitude in this case). 
Only a few faint stars do not have any 
visual magnitude at all.\\

\item {\it Colour index B$-$V}\\
The colour comes from SIMBAD database.\\

\item {\it Photometric flag}\\
The character ":" indicates a large uncertainty in photometry.
A letter "D" indicates a joint magnitude in the case of binaries.
A letter "V" indicates a variable magnitude.\\

\item {\it Effective temperature Teff}\\
The value listed is that which was adopted by the author for the metallicity 
determination in the detailed analysis.
In the previous versions of the Catalogue, we used as effective temperature parameter 
the quantity $\Theta_{eff}={5040 \over Teff}$. We have converted $\Theta_{eff}$ of
the 1991-Edition into Teff, which is now much more widely used. In the preceding
edition, the parameter $\Theta_{eff}$ was presented with two digits of precision.
 The conversion to retrieve Teff has led to an inaccuracy of 
a few tens of degrees for cool stars, about 50 K for solar-like stars, and 
100 to 200 K for hot stars. All the new determinations listed 
in the present edition, give Teff
directly and no correction was necessary for them.\\

\item {\it Logarithm of gravity log\,g}\\
As for the case of Teff, the value of log\,g listed is that which was used by
the author in the spectrum analysis.\\

\item {\it [Fe/H]}\\
 The [Fe/H] value is given with respect to the standard
star listed in column 11.\\

\item {\it Note}\\
The following rare cases are indicated with a letter in this column :\\
- S : [Fe/H] has been obtained with a spectrum of S/N lower than 50\\
- M : several values of [Fe/H] were given in the article (for example from
FeI and FeII lines, or from different instruments), and we give the mean of
them\\
- C : in the absence of any indication of the author, we calculated the         
metallicity with respect to the Sun with $\log \epsilon (Fe)=7.50$\\
- T : combination of cases M and C\\
- D : combination of cases S and C\\

\item {\it Standard }\\
The reference star which was employed in the analysis. \\

\item {\it Reference}\\
Bibliographical reference with respective number to be found in File 3.\\

   \end{enumerate}

In this Edition, as in the previous 1991-Edition, the microturbulence velocity has been
omitted because not all authors use the same definition for it. People interested
in the value of the microturbulence must consult individual references.

File 1 includes 4716 determinations of [Fe/H] for 2497 different stars. File 2 includes 1230
determinations of [Fe/H] for 751 stars in 84 associations, clusters or nearby galaxies.

File 3 gives the list of the 700 bibliographical references (130 more than in the previous
version) sorted in chronological order. We have adopted for all the references the journal abbreviations
of the NASA-ADS Abstract Service.

Several mistakes have been corrected in the previous edition and in this latest version, 
but the Catalogue undoubtedly still contains incorrect data and misprints. We 
encourage users to bring to our attention any error they find. The Catalogue is
kept at the CDS and available by the usual methods. 
It will be updated and corrected
regularly. 


\section{Some remarks on the parameters listed in the Catalogue}

The authors of this Edition decided to give to the Catalogue a more spectroscopic aspect
than in the previous versions, keeping only
as photometric parameters, the visual magnitude V and the colour index B$-$V. As  a great work has been
done by Pr. Hauck and Drs Mermilliod to bring the GCPD up to date, 
it would have been superfluous
to repeat the same information in this spectroscopic Catalogue. Then, why has
a Catalogue of atmospheric parameters of stars been called a "Catalogue of [Fe/H] determinations" ?
Because the understanding of the chemical composition of stars, interstellar matter, and
galaxies has become one of the central issues of modern astronomy. Therefore we wanted to emphasize
the importance of the abundance parameter [Fe/H].

As was already mentionned in Sect. 2, the atmospheric parameters listed in the Catalogue are
deduced from
models generally computed with the assumption
of LTE. But not all the time. The reader is urged to consult the individual references
for more details on the techniques used in each analysis.

The colour index B$-$V is a convenient photometric parameter 
because it has been  determined
for a very large number of stars (in File 1, 98\% of the stars have a colour index B$-$V). 
This index is commonly used as a temperature indicator in galactic studies in which
a large number of very faint stars are considered. The plot in Figure 1 shows the distribution of the 
field stars of the Catalogue (File 1) in the
plane (B$-$V,Teff) for stars ranging between $3000^oK$ and $8000^oK$. 
There is a clear trend of
Teff with respect to B$-$V, but the dispersion about the
correlation is very high.
The relation between 
Teff and B$-$V is particularly
uncertain around B$-$V=0.5 where the corresponding Teff spans more than $1000^oK$.
The dispersion in this relationship is undoubtedly heavily influenced by the 
sensitivity of B$-$V to the metallicity of the star, showing the limitation
of B$-$V as a temperature indicator.


\begin{figure*}
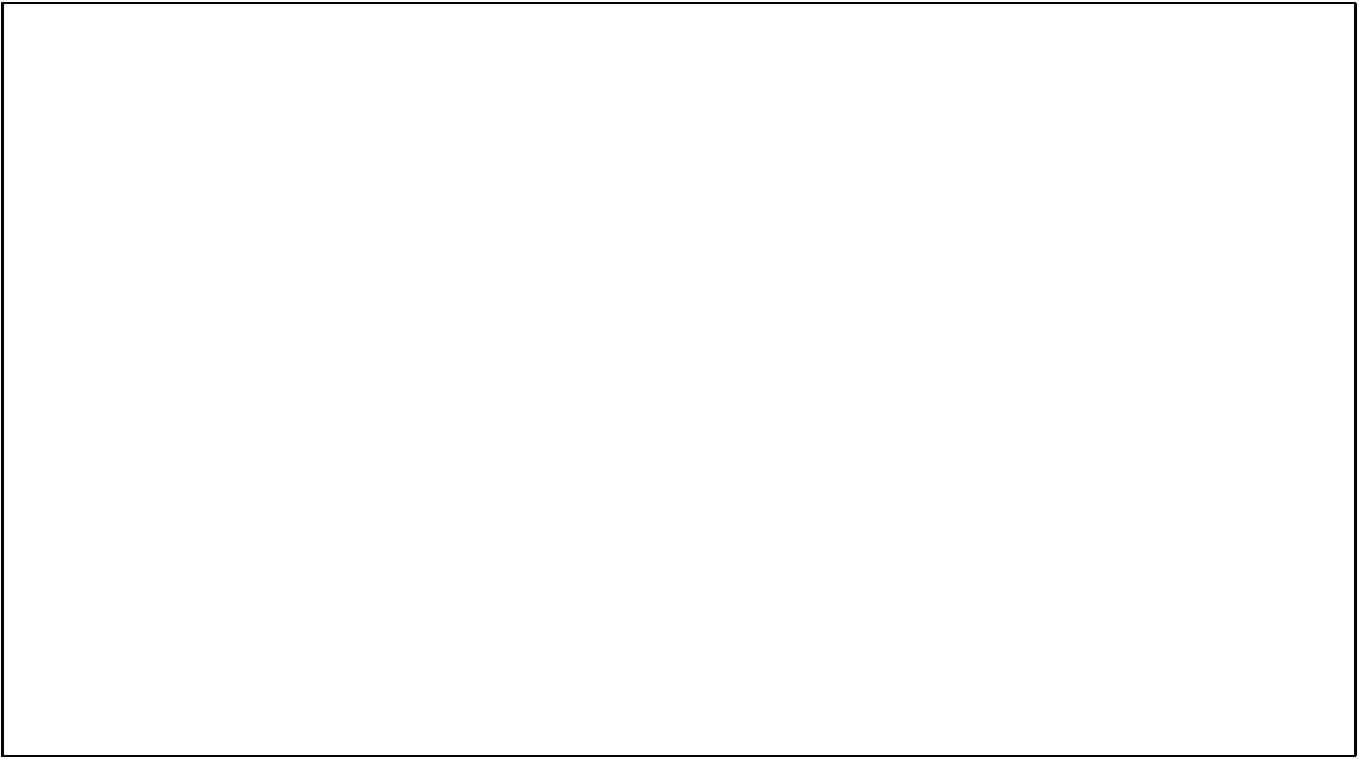

\picplace{10.0cm}
\caption[]{Teff plotted against B$-$V colour for field stars (File 1) of the
Catalogue }
\end{figure*}


In most cases, the Sun is used as the reference star in the analyses. However, for
early type stars, peculiar Ap and Am stars, yellow giants etc ... other well
analysed stars like $\epsilon$ Vir, $\alpha$ Boo or $o$ Peg, are also
used as reference.
In some cases, the
authors have not observed their reference stars with the same
observational and optical equipment as their program stars. Also 
they have not always analysed their program and reference stars 
using the same model
atmospheres and the same physical and atomic parameters. As a result, 
even for most recent analyses, the abundance determinations from different
authors may show significant disagreement in spite of the excellent quality of the 
observations. Instead of taking the mean of the abundances listed for a star with
many determinations, the reader is encouraged to carefully examine the details
of individual analyses to understand the differences. An illustration of this problem
can be given with the well-known star $\epsilon$ Vir (HD 113226). This star is frequently
used as a comparison star in detailed analyses. The Catalogue gives 15 values of [Fe/H] with
respect to the Sun for
$\epsilon$ Vir, ranging between -0.10 dex and +0.21 dex. What value should be taken as the 
true metallicity of $\epsilon$ Vir ? Due to such disagreements, we did not 
attempt to standardize the [Fe/H] values of the Catalogue to the solar one.

 We have been requested several times, between the successive editions, to 
include 
an average value of the different determinations of [Fe/H] for each star. But also
for this Edition we have upheld our initial decision 
to publish the original values from
the literature, the mean value being subjected to change at each edition, 
and a straight mean between
values of unequal quality being physically unjustified. A weighted mean 
would have involved a number
of subjective judgements and the development of some scheme for compensating
for systematic differences and normalizations between individual 
investigations.
For those who want to use a mean anyway,
without making an "educated" weighting, we recommend keeping 
only recent determinations
obtained with high S/N ratios, from spectroscopy with Reticon or CCD detectors. However it must be
noted that [Fe/H] determinations are also affected by the adopted effective temperatures, gravities
and microturbulent velocities, and that a stellar metal abundance can be in error, even if the
observations are of excellent quality.


\section{Some remarks on the stellar content of the Catalogue}

The sample of stars of this new Edition of the Catalogue has conserved several biases, and 
cannot be considered as
representative of the stellar content of the solar neighbourhood.
Figure 2 represents the distribution of
observations listed in File 1 in the plane (Teff, log\,g).
Some parts of this diagramme have a higher density of observations than in an
usual HR diagramme. This correspond probably to the personal interest
of astronomers in problems concerning Am stars $(\sim 6400^oK)$, F dwarfs and subgiants $(\sim 6050^oK)$, 
early G solar type stars $(\sim 5750^oK)$, early G giants $(\sim 4950^oK)$,
late G and K giants $(\sim 4550^oK)$.   
One can note also a lack of 
G and K dwarfs. The principal reason is that these stars,
being intrinsically faint, are more difficult to observe at high resolution and high S/N 
than the giants of the same Teff.
It is a pity for the study of stellar evolution, because it is well known that 
 the 
full span of ages is
still present
among low mass G and K and later type stars. M star spectra are very complicated
and are difficult to analyse in detail, and in this Edition there is still a very low number
of them.\\


\begin{figure*}
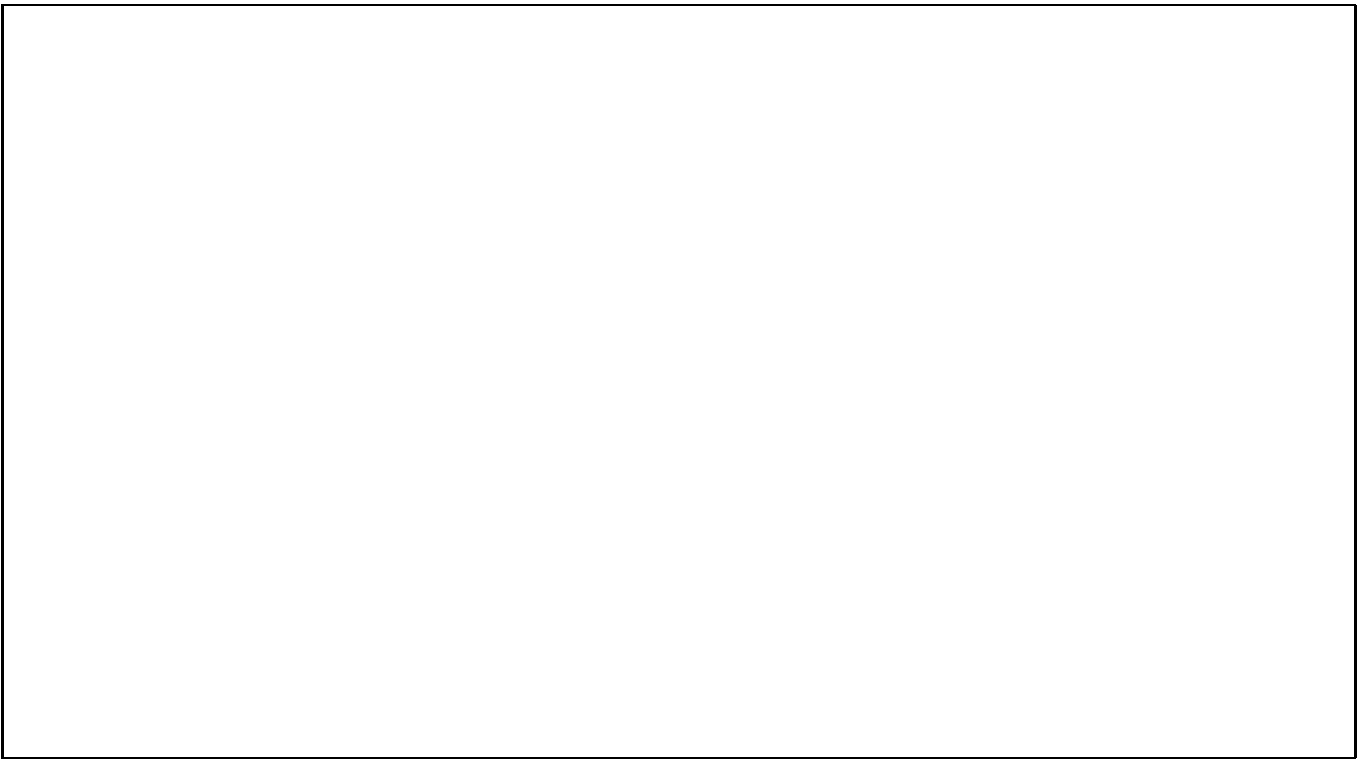

\picplace{10.0cm}
\caption[]{log\,g plotted against Teff for all the entries of File 1
in the temperature range $[8500^oK,3500^oK]$.}
\end{figure*}


In Figure 3, we present an histogram of the 4716 [Fe/H] determinations
given in File 1.  The most frequent metallicity is around -0.15 dex.  
This value is of the same order as the mean metallicity of the solar
neighborhood
found from $ubvy$ photometry by Eggen (1978).  It is important 
to remember that the sample listed in the Catalogue is strongly
biased and no
statistical conclusion can be given, but one can note that the solar metallicity is on            
the metal rich part
of the histogram, which is in the same sense than Eggen's conclusion : the 
Sun seems to have a higher
metallicity than the majority of stars in the solar neighbourhood.

Non-solar metallicities, as shown in Figure 3, may reflect very different
astrophysical phenomena.  Many stellar metallicities reflect the 
interstellar medium from which the star formed, and whose chemical 
composition is appreciably different from the Sun.  This interpretation
applies to most solar and lower mass stars, and it is just these, from 
late F or early G type, that span the full range of stellar ages
present in galaxies.  Abundances of these stars interest experts in 
chemical evolution, age and population effects in field stars and cluster 
stars.  

The peculiar [Fe/H] values found in more massive stars reflect not
the original abundances, but the alteration by physical
processes (selective mass-loss, radiative diffusion, modified or not
by magnetic fields, nuclear burning induced by convection and turbulent mixing).
These stars are studied by experts of the physical structure of stars.


\begin{figure}
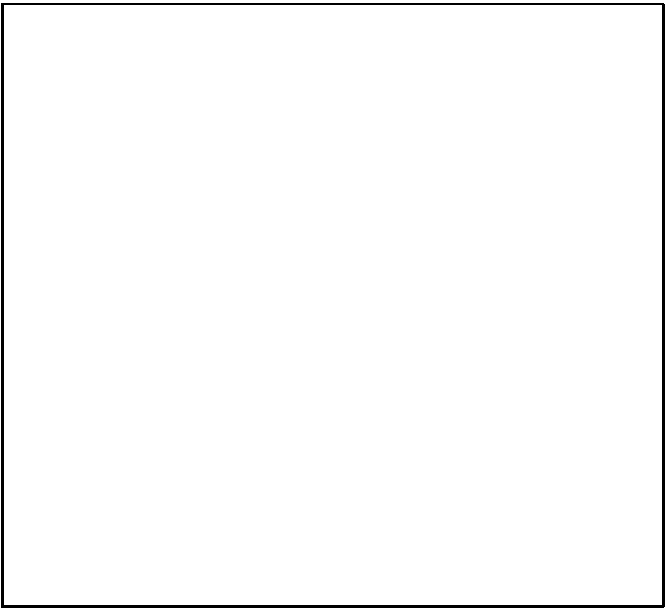

\picplace{8.0cm}
\caption[]{Distribution of the 4716 [Fe/H] determinations in the File 1}
\end{figure}


 An interesting feature of the histogram in Figure 3 is the fraction
of deficient metallicities which is much higher than the true proportion of metal-poor
stars in the solar neighbourhood. 
The comparison of this histogram with the same one published
in the 1991-Edition shows that the number of deficient stars observed during the last 5 years
has considerably increased. A 
large number of nearby metal deficient stars have been observed also
 in several surveys of halo 
and thick disk stars, 
conducted either with low resolution spectroscopy (Beers et al. 1992), 
or from high resolution spectroscopy
at low S/N ratio, from high proper motion lists (Carney et al. 1994). It 
must be mentioned that
over 3000 determinations of [Fe/H] have been obtained by these authors, but only [Fe/H] obtained
by detailed analyses triggered by those lists are quoted in our Catalogue. 
The interest in the first generation of stars will continue to grow
in the next years because these stars provide clues about the chemical 
evolution of the Galaxy, and on nucleosynthesis processes in general. 
The problem is that halo and thick disk stars are faint and difficult to
observe at high dispersion. Figure 4 shows the histogram of the visual
magnitude of the field stars of File 1. It can be seen that the proportion of
stars fainter than V=10 which have been observed at high resolution, is quite low. The dotted 
histogram represents the stars of File 1 which have [Fe/H] $\leq$ -0.8. These
stars are the faintest of the sample.


\begin{figure}
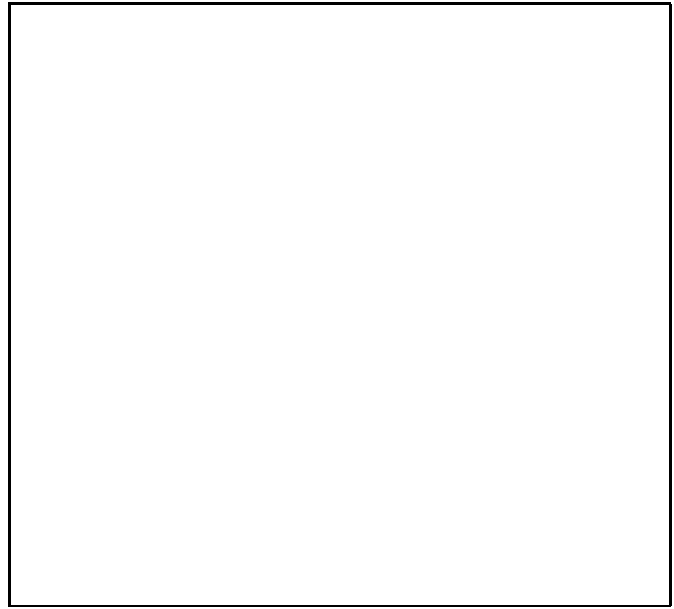

\picplace{8.0cm}
\caption[]{Distribution of the visual magnitude V for the nearly 2500 stars of File 1. The dotted
histogram represent the metal deficient stars ([Fe/H] $\leq$ -0.8)} 
\end{figure}


The problem of the limiting magnitude of high resolution spectroscopy is 
particularly important for clusters and the Magellanic Clouds. 
Except for a few nearby
clusters, the 84 systems listed in File 2 are  very distant and the most
frequent magnitude of the stars listed in this table is V=12.5. Another
consequence of the large distance of clusters is that the cool dwarfs are poorly
represented in File 2 by comparison to giants. Also, only 43\% of the stars of the sample 
have a spectral type, by contrast to File 1 where 96\% of the field stars have
a spectral type. The wide availability of 10-m class telescopes will 
soon increase greatly 
the limiting magnitude of high resolution spectroscopy and this will have
an important impact on the understanding of the chemical properties of the
globular clusters and nearby galaxies.

\section{Discussion and conclusion}

The Catalogue of [Fe/H] determinations provides a database for various
topics related to chemistry in the stars or in the galaxies. This Catalogue 
has been foreseen for two 
kinds of users. The first, often photometrists, are those interested in the
Catalogue because they are specialists in the chemical evolution of
the Galaxy, or of other galaxies (see for instance the calibrations
performed by Golay et al. 1977a, 1977b, 1978).

The second kind of user of the Catalogue, and probably the more numerous, is
working in high resolution high S/N spectroscopy. These astronomers use the
Catalogue in preparing their observing programs, in verifying whether some of their
program stars have high resolution, high S/N observations, and in studying what kind of
literature already exists in connection with their observing program. And, once having
reduced their spectra, the Catalogue can be useful in helping the authors to interpret
the results.

We want to conclude by saying that many of the stars contained in the Catalogue have
been included in different programs for the Hipparcos mission, particularly those
concerning the observational study of galactic structure and chemical evolution
of the solar neighbourhood. The imminent availability of precise distances and
proper motions for those stars, together with good atmospheric parameters, will
give the opportunity to study the correlation between kinematics and metallicity
with excellent observational material.

\begin{acknowledgements}
First of all, we want to thank Pr. Hauck for his constant help during the 
previous editions of the Catalogue. Then, our thanks go to the successive
collaborators of the Catalogue, who 
spent energy and time working for it. We are also very thankful to those
of our colleagues who directed our attention to newly published analyses, and 
to others who provided their data in
electronic form or who sent us lists of errors contained in the Catalogue.
Many thanks  to Monique and Fran\c cois Spite for sending remarks and corrections all
along these years.
We made extensive use of the SIMBAD and NASA-ADS databases, and we are 
extremely grateful to the staff of these two services for maintaining such
valuable resources. And at the end, we want to thank again the members of the
Centre des Donn\'ees Stellaires in Strasbourg for keeping up to date the 
computerized version of this Catalogue.
\end{acknowledgements}

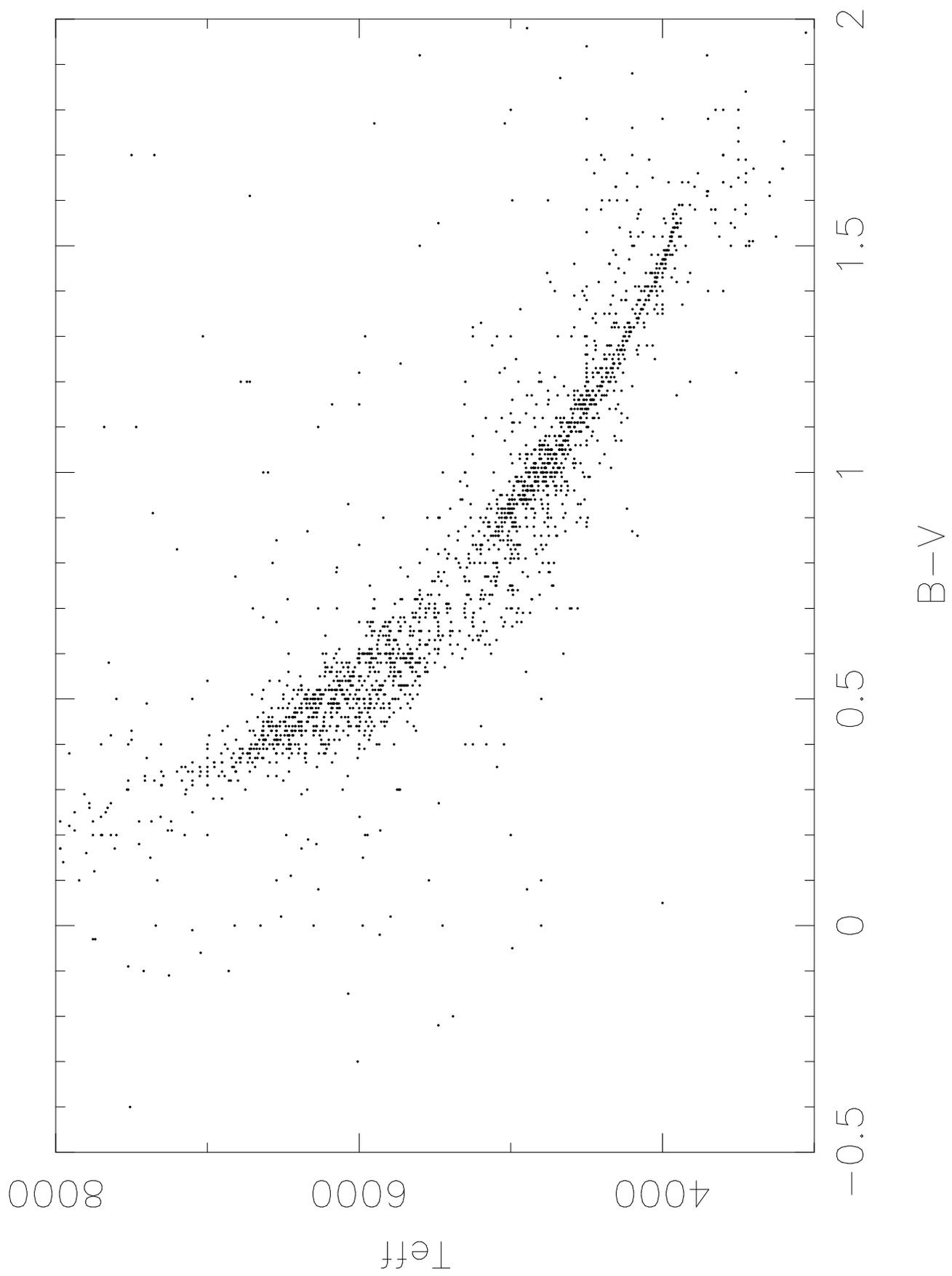
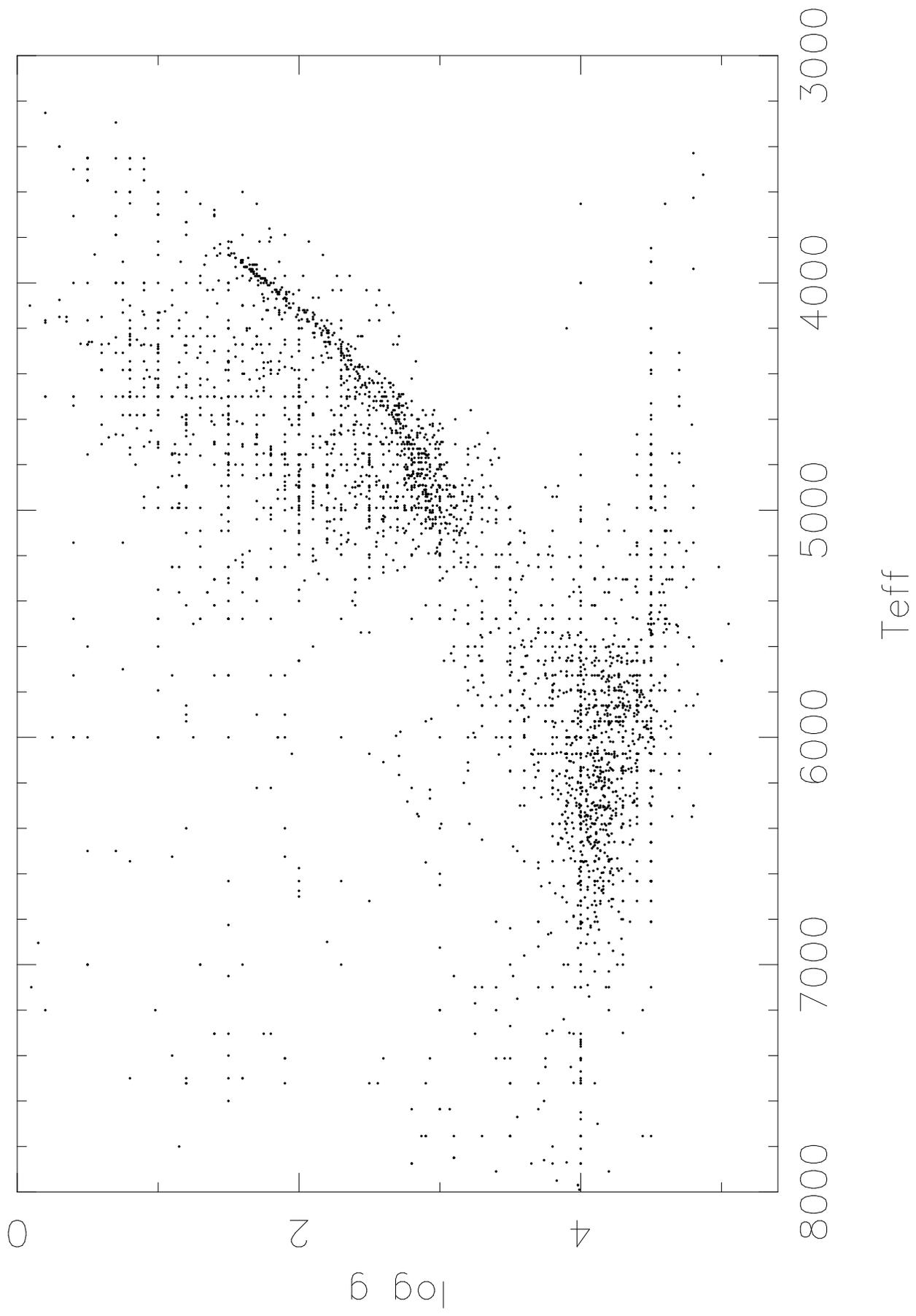
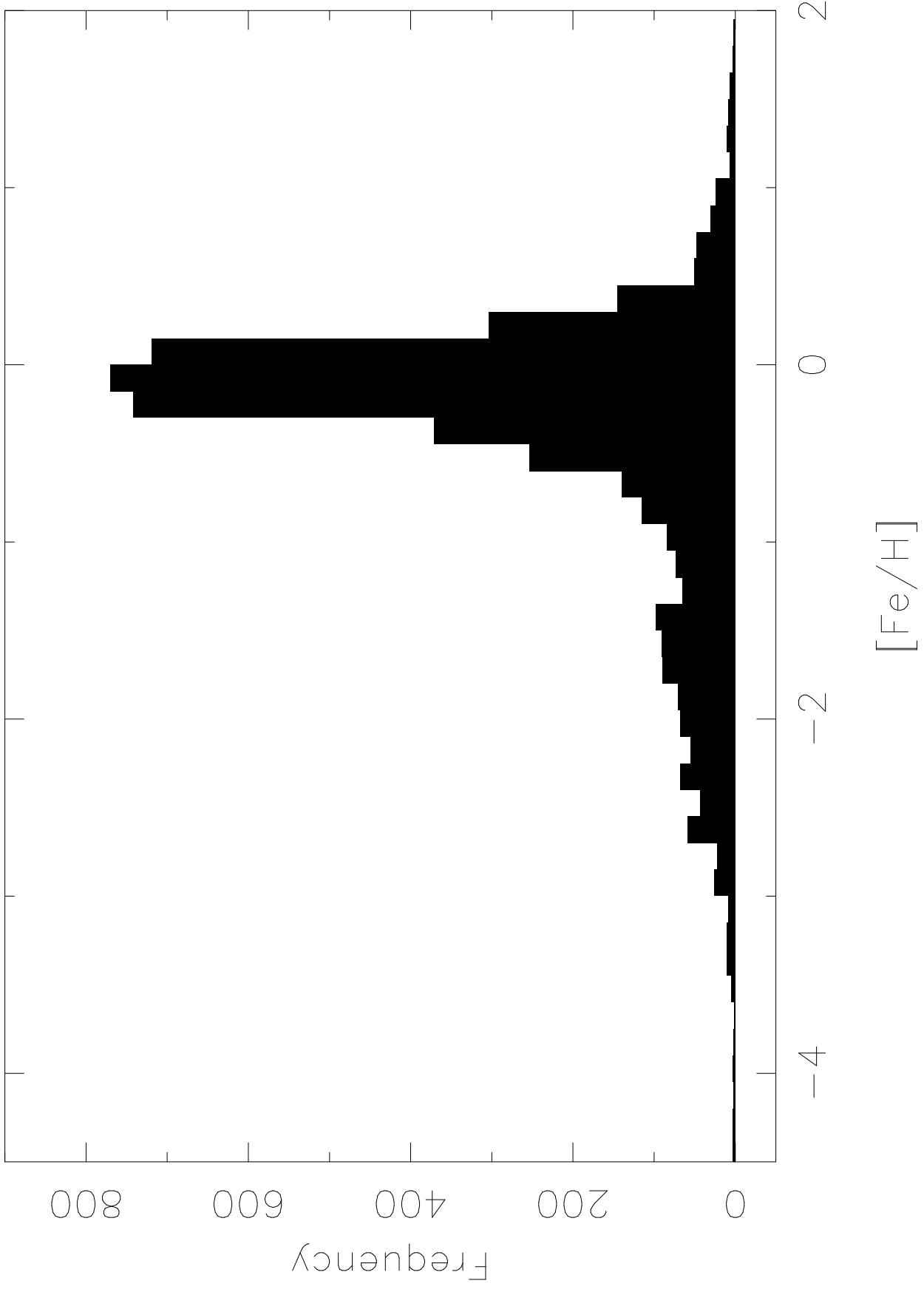
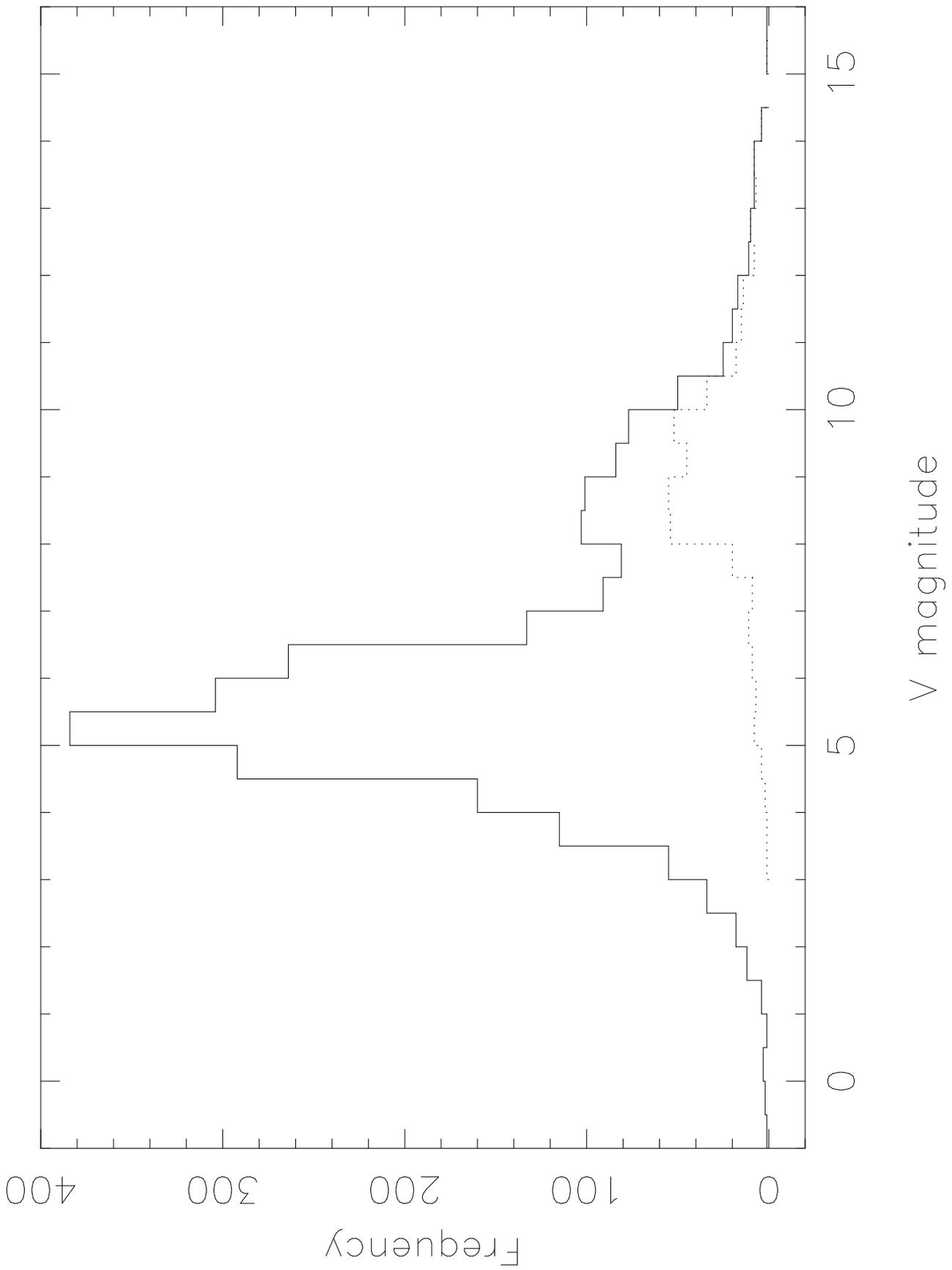
\end{document}